\documentstyle[sprocl]{article}

\def\Journal#1#2#3#4{{#1} {\bf #2}, #3 (#4)}

\def\mE{\mbox{\bf E}}
\def\mH{\mbox{\bf H}}

\def\dmunu{_{\mu\nu}}
\def\pa{\partial}

\def\ppnp{{\it Prog. Part. Nucl. Phys.}\ }

\def\etal{{\it et al.}}
\def\ie{{\it i.e. }}

\def\ie{{\it i.e. }}

\def\etal{{\it et al.}}

\def\be{\begin{equation}}
\def\ee{\end{equation}}
\def\bea{\begin{eqnarray}}
\def\eea{\end{eqnarray}}

\begin{document}

\title{A NEW APPROACH TO QUASARS AND TWINS BY COSMOLOGICAL WAVEGUIDES}

\author{S. CAPOZZIELLO, G. IOVANE}

\address{Dipartimento di Scienze Fisiche ``E.R. Caianiello'',
Universit\`a di Salerno,\\I-84081 Baronissi, Salerno,
ITALY\\E-MAIL: capozziello,geriov@vaxsa.csied.unisa.it}

\maketitle\abstracts{The huge luminosity of quasars and
"twin''images of them could be explained using filamentary or
planar cosmological structures acting as waveguides. We describe
the gravitational waveguide theory and then we discuss possible
realizations in cosmology.}

Gravitational lensing is today a fundamental tool to investigate
the large scale structures of the universe and to  test
cosmological models. One of the most interesting characteristics
of it is that it acts on all scales and provides a great amount of
cosmological and astrophysical applications. Particularly
promising are the multiply macro--imaged quasars whose lensing
galaxy should have a large optical depth for lensing effects (at
least 20 objects of this kind  have been identified till now). The
gravitational lensing pheomena can be explained through the action
of a weak gravitational field on the light rays. In optics,
however, there exist other types of devices, like optical fibers
and waveguides which use the same deflection phenomena. The
analogy with the action of a gravitational field onto light rays
may be extended considering also these other structures. In other
words, it is possible to suppose the existence of gravitational
waveguides  considering large scale structures as filaments,
clusters and groups of galaxies \cite{capozziello}. If this
happens, for example, a filament of galaxies could preserve the
total luminosity of a far source as a quasar. The discussion of
gravitational waveguide properties can be done starting from the
electromagnetic field theory in a gravitational field described by
the metric tensor $g\dmunu$ \cite{ehlers}. A waveguide solution is
obtained by reducing the Maxwell equations in a medium to the
Helmholtz scalar equations for the fields $\mE$ and $\mH$
\cite{noi}. For some arbitrary monochromatic component of the
electric field, we have
\be
\label{18} \frac{\pa^{2}E}{\pa z^{2}}+\frac{\pa^{2}E}{\pa x^{2}}
+\frac{\pa^{2}E}{\pa y^{2}}+k^{2}n^{2}({\bf r})E=0\,, \ee where
$k$ is the wave number. The coordinate $z$, in Eq.(\ref{18}), is
the longitudinal one, and it can measure the space distance along
the structure produced by a mass distribution with an optical
axis. Such a coordinate may also correspond to a distance along
the light path inside a planar matter--energy distribution in some
regions of the universe. Let us consider a solution of the form
\be
\label{1}
E=n_{0}^{-1/2}\Psi\exp\left(ik\int^{z}n_{0}(z')dz'\right)\;;
 \;\;\;\;\;n_{0}\equiv n(0,0,z)\,,
\ee where $\Psi(x,y,z)$ is a slowly varying spatial amplitude
along the $z$ axis, and $\exp(iknz)$ is a rapidly oscillating
phase factor. It is clear that the beam propagation  is along the
$z$ axis. We rewrite Eq.(\ref{18}),  neglecting second order
derivative in longitudinal coordinate $z$, and obtain a
Schr\"odinger--like equation for $\Psi$:
\be
\label{2} i\lambda\frac{\pa \Psi}{\pa \xi}=-\frac{\lambda^{2}}{2}
\left(\frac{\pa^{2}\Psi}{\pa x^{2}}+\frac{\pa^{2}\Psi}{\pa
y^{2}}\right)
+\frac{1}{2}\left[n_{0}^{2}(z)-n^{2}(x,y,z)\right]\Psi\,, \ee
where $\lambda$ is the electromagnetic radiation wavelength and we
adopt the new variable $\xi=\int^{z}dz'/n_{0}(z'),$ normalized
with respect to the refraction index. If one has the distribution
of the matter in the form of cylinder with a constant (dust)
density $~\rho_{0} ~$,  the gravitational potential inside
 has a  parabolic  profile providing a waveguide effect
for electromagnetic radiation analogous to  that realized in
optical fibers. In this case, Schr\"odinger--like equation is that
of two--dimensional quantum harmonic oscillator for which the mode
solutions exist in the form of Gauss--Hermite polynomials. As a
side remark, it is interesting to stress that, considering again
Eq.(\ref{2}), the term in square brackets in the rhs plays the
role of the potential in a usual Schr\"odinger equation; the role
of Planck constant is now assumed by $\lambda$. Since the
refraction index can be expressed in terms of the Newtonian
potential when we consider the propagation of light in a
gravitational field, we can write the  potential in (\ref{2}) as
\be
\label{pot} U({\bf r})=\frac{2}{c^{2}}[\Phi(x,y,z)-\Phi(0,0,z)]\,.
\ee The waveguide effect depends specifically on the shape of
potential (\ref{pot}): for example,  the radiation from a remote
source does not attenuate if $U\sim r^{2}$; this situation is
realized supposing a "filamentary'' or a "planar'' mass
distribution with constant density $\rho$. Due to the Poisson
equation, the potential inside the filament is a quadratic
function of the transverse coordinates, that is of
$r=\sqrt{x^{2}+y^{2}}$ in the case of the filament and of $r=x$ in
the case of the planar structure (obviously the light propagates
in the "remaining'' coordinates: $z$ for the filament, $z,y$ for
the plane). Then, if the radiation, travelling from some source,
undergoes a waveguide effect, it does not attenuate like $1/R^{2}$
as usual, but it is, in some sense conserved; this fact means that
the source brightness will turn out to be much stronger than the
brightness of analogous objects located at the same distance (\ie
at the same redshift $Z$) and the apparent energy released by the
source will be anomalously large. To fix the ideas, let us
estimate how the  electric field (\ref{1}) propagates into an
ideal  filament whose internal potential is
\be
\label{internal} U(r)=\frac{1}{2}\omega^{2}r^{2}\,,\;\;\;\;\;\;
\omega^{2}=\frac{4\pi G \rho}{c^{2}} \ee where $\rho$ is constant
and $G$ is the Newton constant.
 A spherical wave from a source,
$E=(1/R)\exp(ikR),$ can be represented in the paraxial
approximation as
\be
\label{4} E(z,r)=\frac{1}{z}\exp\left(ikz+\frac{ikr^{2}}{2z}-
\frac{r^{2}}{2z^{2}}\right)\,, \ee where we are using the
expansion
\be
\label{5} R=\left(z^{2}+r^{2}\right)^{1/2}\approx
z\left(1+\frac{r^{2}}{2z^{2}}\right)\,,\;\;\;\;r\ll z\,. \ee It is
realistic to  assume $n_{0}\simeq 1$ so that $\xi=z$. Let us
consider now that the starting point of the filament of length $L$
is at a distance $l$ from a source shifted by a distance $a$ from
the filament axis in the $x$ direction. The amplitude  $\Psi$ of
the field $E$, entering the waveguide is
\be
\label{6} \Psi_{in}=\frac{1}{l}\exp
\left[\frac{ikl-1}{2l^{2}}\left((x-a)^{2}+y^{2}\right)\right]\,,
\ee and so we have $ R=\left(l^{2}+y^{2}+(x-a)^{2}\right)^{1/2}.$
We can calculate the amplitude of the field at the exit of the
filament by the equation
\be
\label{8} \Psi_{f}(x,y,l+L)= \int dx_{1}dy_{1}{\cal
G}(x,y,l+L,x_{1},y_{1},l)\Psi_{in}(x_{1},y_{1},l)\,, \ee where
${\cal G}$ is the Green function of Eq.(\ref{2}) which is the
propagator of the harmonic oscillator. The integral (\ref{8}) is
Gaussian and it can be exactly evaluated. It is interesting in two
limits. If $\omega l\ll 1$, we have
\be
\label{112} \Psi_{f}=\frac{1}{i\lambda}\exp
\left\{-\frac{l+i\lambda}{2\lambda^{2}l}\left[(x+a)^{2}+y^{2}\right]\right\}\,,
\ee which means that the radiation emerging from a point with
coordinate $(a,0,0)$ is focused near a point with coordinates
$(-a,0,l+L)$ (that is the radius has to be of the order of the
wavelength). This means that, when the beam from an extended
source is focused inside the waveguide in such a way that, at a
distance $L$, an inverted image of the source is formed, having
the very same geometrical dimensions of the source. The waveguide
``draws" the source closer to the observer since, if the true
distance of the observer from the source is $R$, its image
brightness will correspond to that of a similar source at the
closer distance $R_{eff}=R-l-L\,.$ If we do not have $\omega l\ll
1$, we get (neglecting the term $i\lambda/l$ compared with unity)
\be
\label{13} \Psi_{f}=\frac{\sqrt{1+(\omega l)^{2}}}{i\lambda} \exp
\left\{-\frac{1+(\omega l)^{2}}{2\lambda^{2}}
\left[y^{2}+\left(x+\frac{a}{\sqrt{1+(\omega l)^{2}}}\right)^{2}
\right]\right\}\,, \ee from which, in general, the size of the
image is decreased by a factor $\sqrt{1+(\omega l)^{2}}$. The
amplitude increases by the same factor, so that the brightness is
$(R/R_{eff})$ times larger. In the opposite limit $\omega l\gg 1$,
we have
 $L\simeq \pi/\omega$, that is
the shortest focal length of the waveguide is
\be
\label{foc} L_{foc}=\sqrt{\frac{\pi c^{2}}{4G\rho}}\,, \ee which
is the length of focusing of the initial beam of light trapped by
the gravitational waveguide. All this arguments apply if the
waveguide has (at least roughly) a cylindrical geometry. The
theory of planar waveguide is similar but we have to consider only
$x$ as transverse dimension and not also $y$. The cosmological
feasibility of a waveguide depends on the geometrical dimensions
of the structures, on the connected densities and on the limits of
applicability of the above idealized scheme. Due to very large
scale sizes of the structure (we give an extimation below),
 the electromagnetic
radiation deflection by the gravitational waveguide occurs and, in
principle, it may be observed. We will mention, for example, a
possibility of brilliancy magnification of the long distanced
objects (like quasars) with large redshifts as consequence of the
waveguiding structure existence between the object and the
observer. This effect exists also for a gravitational lens located
between the object and the observer, but the long gravitational
waveguide may give huge magnification, since the radiation
propagates along the waveguide with functional dependence of the
intensity on the distance which does not decrease as $~\sim
1/R^2~$, characteristic for free propagation. The gravitational
lens, being a compact object, collects much less light by
deflecting the rays to the observer than the gravitational
waveguide structure transporting to the direction of observer all
trapped energy (of course, one needs to take into account losses
for scattering and absorbtion). From this point of view, it is
possible that the enormous amount of radiation emitted by quasars
is only seemingly existing. The object may radiate a resonable
amount of energy but the  waveguide structure transmits the energy
in high portion to the observer. In fact, quasars are objects at
very high redshift which appear almost as point sources but have
luminosity that are about one hundred times than that of a giant
elliptical galaxy (quasars have luminosity which range between
$10^{38}\div 10^{41}$ W). The quasars, often, have both emission
and absorption lines in their spectra. The emission spectrum is
thought to be produced in the quasar itself; the absorption
spectrum, in gas clouds that have either been ejected from the
quasar or just happen to lie along the same line of sight. The
brightness of quasars may vary rapidly, within a few days or less.
Thus, the emitting region can be no larger than a few light--days,
i.e. about one hundred astronomic units. This fact excludes that
quasars could be galaxies (also if most astronomers think that
quasars are extremely active galactic nuclei). The main question
is how to connect this small size with the so high redshift and
luminosity.  A  widely accepted preliminary model forecasts that
most galaxies contain a compact central  nucleus with mass
$10^{7}\div 10^{9}$ M$_{\odot}$ and diameter $< 1 $ pc. Such a
nucleus may release an amount of energy exceeding the power of all
the rest of the galaxy. However the mechanism is still  not very
well known. Some people suppose that it is connected to the
releasing of gravitational energy due to the interactions of
internal components of quasars. This mechanism is extremely more
efficient than the releasing of energy during the ordinary nuclear
reactions. The necessary gravitational  energy could be produced,
for example, as consequence of the falling of gas in a very deep
potential well as that connected with a very massive black hole.
Only with this assumption, it is possible to justify a huge
luminosity, a cosmological
 redshift and a small size for the quasars.
An alternative explaination could come from waveguiding effects.
As we have discussed, if light travels within a filamentary or a
planar structure, whose Newtonian gravitational potential is
quadratic in the transverse coordinates, the radiation is not
attenuated, moreover the source brightness is stronger than the
brightness of analogous object located at the same distance (that
is at the same redshift). In other words, if the light of a quasar
undergoes a waveguiding effect, due to some structure along the
path between it and us, the apparent energy released by the source
will be anomalously large, as the object were at a distance given
by $R_{eff}$. Furthermore, if the approximation $\omega l\ll 1$
does not hold,  the dimensions of resulting image would be
decreased by a factor $\sqrt{1+(\omega l)^{2}}$ while the
brightness would be $(R/R_{eff})^{2}$, larger, then explaining how
it is possible to obtain so large emission by such (apparently)
small  objects.  The existence of a waveguiding effect may
prevents to take into consideration exotic mechanism in order to
produce huge amounts of energy (as the existence of a massive
black hole inside a galactic core) and it may justify why it is
possible to observe so distant objects of small geometrical size.
Another effect concerning the quasars may be directly connected
with multiple images in lensing. The waveguide effect does not
disappear if the axis of the filament or if the guide plane is
bent smoothly in space. As in the case of gravitational lenses, we
can observe "twin'' images if part of the radiation comes to the
observer directly from the source, and another part is captured by
the bent waveguide. The "virtual'' image can then turn out to be
brighter than the "real" one (in this case we may deal with
"brothers'' rather than "twins'' since parameters like, spectra,
emission periods and chemical compositions are similar but the
brightnesses are different). Furthermore, such a bending in
waveguide could explain large angular separations among the images
of the same object which cannot be explained by the current lens
models (pointlike lens, thin lens and so on). A further remark has
to be done on the densities of waveguide structures which allow
observable effects. Considering Eq.(\ref{foc}) and introducing
into it the critical density of the universe $\rho_{c}\sim
10^{-29}$ g/cm$^{3}$ (that is the value for which the density
parameter is $\Omega=1$), we obtain $L_{foc}\sim 5\times 10^{4}$
Mpc which is an order of magnitude larger than the observable
universe and it is completely unrealistic. On the contrary,
considering a  typical galactic density as $\rho\sim 10^{-24}$
g/cm$^{3}$, we obtain $L_{foc}\sim 100$ Mpc, which is a typical
size of large scale structure. However, an important issue has to
be taken into account: the absorption and the scattering of light
by
 the matter inside the filament or the planar structure increase with
density and, at certain critical value the waveguide effect, can
be invalidated. It can be shown that no restriction exists if the
radio band
 and a thickness of the structure $r>10^{14}$cm are considered
 \cite{capozziello}.
In such a case, the relative density change between the background
and the structure density is valid till $\delta \rho/\rho\leq 1$.
By such hypotheses, practically all the observed large scale
structures can result as candidates for waveguide effects.

Finally, it is possible to construct a simulation where, given
random distributions of quasars and waveguides, we reproduce the
observed luminosity function of quasar distribution
 explaining its anomalies \cite{noi}.
In particular, the observed excess (resulting in a duble peak) of
the number of objects against the redshift $Z$, is easily
explained by the waveguide hypothesis Furthermore, the waveguide
model does not concern only electromagnetic radiation: a waveguide
effect could be observed also for streams of neutrinos or for
gravitational waves which interact with large scales structures
\cite{noi}.

\vfill
\end{document}